\documentclass[a4paper,11pt]{article}
\usepackage{pos}

\usepackage{amsthm}
\usepackage{mathtools}
\usepackage{graphicx}
\usepackage{amsmath}
\usepackage{amsfonts}
\usepackage{mathrsfs}
\usepackage{bbm}
\usepackage{bm}
\usepackage[T1]{fontenc}

\def\ds{\stackrel{\star}{,}}

\def\ds{\stackrel{\star}{,}}

\newcommand{\dwedge}{\curlywedge}

\numberwithin{equation}{section}
\catcode`@=12

\def\={\ =\ }
\def\dd{{\rm d}}

\newcommand{\Tr}[1]{\:{\rm Tr}#1}

\newcommand{\mbf}[1]{{\boldsymbol {#1} }}

\def\lact{{\textrm{\tiny L}}}
\def\ract{{\textrm{\tiny R}}}

\def\ii{{\,{\rm i}\,}}

\newcommand{\bbr}{\mathbb{R}}

\newcommand{\bbc}{\mathbb{C}}

\newcommand{\calf}{\mathcal{F}}

\newcommand{\cali}{\mathcal{I}}

\def\alg{{\mathcal A}}

\newcommand{\bbz}{{\mathbb Z}}

  \definecolor{violet}{rgb}{.7,0,1}

  \definecolor{dgreen}{rgb}{.2,.5,.1}

  \definecolor{test}{rgb}{.5,1,0}

\newcommand{\CCF}{\mathscr{F}}

\newcommand{\CL}{\mathcal{L}}

\newcommand{\CCR}{\mathscr{R}}

\newcommand{\CV}{\mathcal{V}}

\newcommand{\frg}{\mathfrak{g}}				

\newcommand{\fru}{\mathfrak{u}}

\newcommand{\str}{{\,\mathrm{str}\,}}     		

\newcommand{\aso}{\mathfrak{so}}

\newcommand{\remark}[1]{}     				
\title{The $L_{\infty}$-structure of noncommutative gravity}

\ShortTitle{The $L_{\infty}$-structure of noncommutative gravity}
    
\author*{Richard J. Szabo}

\affiliation{Department of Mathematics, Heriot-Watt University\\
	Colin Maclaurin Building,
	Riccarton, Edinburgh EH14 4AS, U.K.\\ and Maxwell Institute for
	Mathematical Sciences, Edinburgh, U.K.\\
	and Higgs Centre
	for Theoretical Physics, Edinburgh, U.K.}
	
\emailAdd{R.J.Szabo@hw.ac.uk}

\abstract{We summarise recent perspectives on symmetries of noncommutative field theories based on homotopy algebras. We show how these viewpoints naturally lead to a new class of noncommutative field theories which possess braided gauge symmetries, and explain in detail their uses in gravity. We review how these considerations lead to a new theory of noncommutative gravity in four dimensions within the Einstein-Cartan-Palatini formalism.}

\FullConference{Corfu Summer Institute 2021 ``School and Workshops on Elementary Particle Physics and Gravity'' \\
		29 August -- 9 October 2021\\
		Corfu, Greece\\
}

\notes{\note{Preprint:  {\sf EMPG--22--06}}}
	
\begin{document}

\maketitle

\section{Introduction}

In this contribution we briefly review recent developments aimed at understanding gauge symmetries of noncommutative field theories in the framework of homotopy algebras, and in particular $L_\infty$-algebras. For a detailed exposition of the subject and its applications to field theory, see~\cite{Giotopoulos:2021ieg}. Here we shall summarise the applications to gravity, which were developed in a series of papers~\cite{Ciric:2020hfj,Ciric:2020eab,Ciric:2021rhi}. In fact, these investigations were motivated in part by a decade old problem inspired by string theory, that we now briefly recall.

In certain non-geometric flux compactifications of string theory, it is conjectured that the low-energy effective dynamics of closed strings are described by noncommutative or even nonassociative deformations of gravity~\cite{Blumenhagen:2010hj,Lust:2010iy,Blumenhagen:2011ph,Mylonas:2012pg, Blumenhagen:2013zpa} (see~\cite{Szabo:2018hhh,Szabo:2019hhg} for reviews with further references). From a technical standpoint, such effective theories are difficult to construct, because the metric aspects of nonassociative differential geometry are only partially developed~\cite{Mylonas:2013jha,Barnes:2014ksa,Aschieri:2015roa,Barnes:2015uxa, Blumenhagen:2016vpb,Aschieri:2017sug}. In particular, no version of the Einstein-Hilbert action is currently known in this framework. One way around this difficulty is to regard nonassociative gravity as a deformation of `gauge theory', that is, to use the Einstein-Cartan principal bundle formulation, where the corresponding action is the Palatini action~\cite{Barnes:2015uxa,Barnes:2016cjm}.

However, one immediately runs into problems with the naive notion of gauge symmetry. Consider a noncommutative ${ U}(N)$ gauge theory on $\bbr^{d}$ with a nonassociative star-product $\star$ for which the usual de~Rham differential $\dd$ is a derivation, i.e. for which the deformed exterior algebra $(\Omega^\bullet(\bbr^d),\wedge_\star)$ forms a noncommutative differential calculus with $\dd$. A prominent example is the nonassociative star-product that arises via cochain twist quantization of the basic non-geometric $R$-flux deformations of spacetime~\cite{Mylonas:2013jha,Aschieri:2015roa,Aschieri:2017sug}. For a connection one-form $A\in\Omega^1(\bbr^{d},\fru(N))$ and a gauge parameter $\lambda\in\Omega^0(\bbr^d,\fru(N))$, the naive infinitesimal star-gauge transformation would be
\begin{align}\label{eq:naivegauge}
\delta_\lambda^\star A \= \dd\lambda-[\lambda \ds A] \= \dd\lambda-(\lambda\star A-A\star\lambda) \ .
\end{align}
The field strength tensor $F_A^\star\in\Omega^2(\bbr^d,\fru(N))$ is the noncommutative curvature of the connection $A$ defined by
\begin{align}\label{eq:FAstarintro}
F_A^\star\=\dd A + \tfrac12\,[A\ds A]\=\dd A + A\wedge_\star A \ .
\end{align}

Applying \eqref{eq:naivegauge} to the definition \eqref{eq:FAstarintro}, the transformation law of $F_A^\star$ is given by
\begin{align}
\begin{split}
\delta_\lambda^\star F_A^\star & \= -\big(\lambda\star\dd A + (\lambda\star A)\wedge_\star A\big) + \big(\dd A\star \lambda + A\wedge_\star(A\star \lambda)\big) \\
& \qquad + (A\star \lambda)\wedge_\star A - A\wedge_\star(\lambda\star A) \ .
\end{split}
\end{align}
In the associative case, the first line gives exactly $-[\lambda \ds F_A^\star]$, while the second line vanishes. However, for a general nonassociative star-product this is not the case. Thus in general the infinitesimal gauge transformation of $F_A^\star$ does not have the exprected covariant form:
\begin{align}
\delta_\lambda^\star F_A^\star \ \neq \ -[\lambda \ds F_A^\star] \ .
\end{align}
A related problem is that nonassociativity generally obstructs the closure of the infinitesimal gauge transformations \eqref{eq:naivegauge}:
\begin{align}
(\delta^\star_\lambda\,\delta^\star_\rho-\delta^\star_\rho\,\delta^\star_\lambda)A \ \neq \ \delta^\star_{[\lambda \ds \rho]}A \ ,
\end{align}
and indeed the star-commutator $[-\ds-]$ is not generally a Lie bracket on $\Omega^0(\bbr^d,\fru(N))$ as it violates the Jacobi identity when the star-product $\star$ is nonassociative.

A natural setting in which to overcome these problems is the framework of $L_\infty$-algebras. In the physics literature these homotopy algebraic structures first appeared as the natural objects which organise the ``generalized'' gauge symmetries of closed string field theory via an infinite sequence of higher brackets~\cite{Zwiebach:1992ie}. Subsequently it was clarified that any classical perturbative field theory with ``generalized'' gauge symmetries is organised by an $L_\infty$-algebra with finitely-many brackets, due to the duality with the BV--BRST formalism~\cite{Linfty,BVChristian}. We briefly review these ideas in Section~\ref{sec:LinftyCFT}.

The $L_\infty$-structure of general noncommutative and nonassociative gauge theories for star-products which quantize generic almost Poisson structures was partially developed in~\cite{Blumenhagen:2018kwq,Kupriyanov:2019ezf,Kupriyanov:2021cws,Abla:2022wfz} as a means of building the correct notions of gauge symmetry that overcome the problems discussed above, as well as those related to the failure of the de~Rham differential $\dd$ to generally obey the Leibniz rule. These constructions also lead to $L_\infty$-algebras with infinitely-many brackets in general. The drawback of these treatments is that so far they are only understood at the semi-classical level, and hence do not capture the complete structure underlying noncommutative gauge symmetries. Instead, one can reformulate some of these theories via deformations with braided gauge symmetries~\cite{Ciric:2021rhi,Giotopoulos:2021ieg}, which involve the same finite number of brackets as their classical counterparts and are applicable at the full quantized level. This formalism is reviewed in Section~\ref{sec:braided}.

The $L_\infty$-algebras of Einstein-Hilbert gravity are discussed in~\cite{Linfty,Nutzi:2018vkl,Reiterer:2018wcb}. They require perturbing the theory around a flat background and also involve infinitely-many brackets. The $L_\infty$-structure of the {Einstein-Cartan-Palatini (ECP) theory was developed in~\cite{Ciric:2020eab} and it only requires finitely-many brackets.} We review the $L_\infty$-algebra formulation of ECP gravity in Section~\ref{sec:GR}. 

One of the driving motivations behind our development of braided gauge symmetries was the observation that the twisted diffeomorphism symmetry of the usual formulations of noncommutative gravity does not fit naturally into the standard $L_\infty$-algebra picture (if it does at all). We elaborate on this point in Section~\ref{sec:LinftyNCgrav}, which further motivates a deformation of the underlying $L_\infty$-structure itself in order to make it compatible with the twisted diffeomorphisms. We then review the braided $L_\infty$-algebra which organises noncommutative gravity in the example of the four-dimensional ECP theory.

\section{$L_\infty$-algebras of classical field theories}
\label{sec:LinftyCFT}

In this section we give a brief introduction to $L_\infty$-algebras along with their uses in organising the gauge symmetries and dynamics of classical field theories. From a mathematical perspective, their applicability in these contexts comes from their duality with differential graded commutative algebras~\cite{Lada:1992wc,LadaMarkl94}. Our presentation will be mostly informal, in order to highlight the main ideas without too much technical clutter; see~\cite{Linfty,BVChristian,Jurco:2019bvp,Giotopoulos:2021ieg} for more elaborate reviews.

\paragraph{$\mbf{L_\infty}$-algebras.} 
Let $V\= \oplus_{k\in\bbz}\, V_{k}$ be a $\bbz$-graded vector space, with degree shifted graded exterior algebra $\Lambda_V\= \wedge^\bullet(V[1])$ viewed as a free cocommutative coalgebra. An \emph{$L_\infty$-structure} on $V$ is a coderivation $L:\Lambda_V\longrightarrow\Lambda_V$ of degree $|L|\=1$ satisfying $L^2\=0$.
 We can decompose the coderivation $L$ into `components'  $\{\ell_n\}_{n\geq1}$ where $\ell_n:\wedge^n(V[1])\longrightarrow V[1]$ are multilinear maps of degree $|\ell_n|\=1$, or restoring the original grading, maps $\ell_n:\wedge^nV\longrightarrow V$ with degree $|\ell_n|\=2-n$. Then the differential condition $L^2\=0$ leads to an infinite sequence of \emph{homotopy relations}, the first three of which read, together with their meaning, as
\begin{align}\label{eq:Linftyaxioms}
\begin{split}
\ell_1(\ell_1(v))&\=0 \hspace{4.6cm} {(V,\ell_1) \mbox{\sf \ is a cochain complex }} \\[4pt]
\ell_1(\ell_2(v,w))&\=\ell_2(\ell_1(v),w)\pm\ell_2(v,\ell_1(w)) \hspace{1cm} {\ell_1  \mbox{\sf \  is a derivation of } \ell_2} \\[4pt]
\ell_2(v,\ell_2(w,u))+{\rm cyclic}&\=(\ell_1\circ\ell_3\pm\ell_3\circ\ell_1)(v,w,u) \hspace{1cm}  {\mbox{\sf Jacobi identity up to homotopy}}
\end{split}
\end{align}
where the signs depend on the degrees of the elements $v,w,u\in V$. This list continues to include ``higher homotopy Jacobi identities'' involving higher brackets. Thus $L_\infty$-algebras are generalizations of differential graded Lie algebras, which are the special cases where $\ell_n\=0$ for all~$n\geq3$. 

Dualizing this coalgebra picture gives a graded commutative algebra derivation  $Q\=L^*:\Lambda_V^*\longrightarrow\Lambda_V^*$ of degree $|Q|\=1$ satisfying $Q^2\=0$. Thus $L_\infty$-algebras are dual to differential graded commutative algebras.

\paragraph{Quasi-isomorphisms.}
An \emph{$L_\infty$-morphism} from an $L_\infty$-algebra $(V,L)$ to an $L_\infty$-algebra $(V',L')$ is a degree-preserving coalgebra homomorphism $\Psi:\Lambda_V\longrightarrow \Lambda_{V'}$ which intertwines the codifferentials: $\Psi\circ L\=L'\circ\Psi$.
In the `components' $\{\psi_n\}_{n\geq1}$ of $\Psi$, where \ $\psi_n:\wedge^nV\longrightarrow V'$ are maps of degree $|\psi_n|\=1-n$, this condition gives
\begin{align}
\begin{split}
&\hspace{1.1cm} {\psi_1 \mbox{\sf \ is a cochain map: }} \  \psi_1\,\ell_1 \= \ell_1'\,\psi_1  \\[4pt]
&\psi_1(\ell_2(v,w))-\ell_2'(\psi_1(v),\psi_1(w)) \ {\mbox{\sf \ is a homotopy in } \psi_2}
\end{split}
\end{align}
plus cumbersome higher relations. Thus $L_\infty$-morphisms generalize homomorphisms of differential graded Lie algebras. 

An $L_\infty$-morphism is an {\it $L_\infty$-quasi-isomorphism} if the induced map on cohomology $\psi_{1*}:H^\bullet(V,\ell_1)\longrightarrow H^\bullet(V',\ell_1')$ is an isomorphism. Quasi-isomorphism is an equivalence relation on $L_\infty$-algebras, contrary to quasi-isomorphism on the smaller category of differential graded Lie algebras.

\paragraph{Cyclic $\mbf{L_\infty}$-algebras.}
A \emph{cyclic structure} on an $L_\infty$-algebra $(V,L)$ is a graded inner product  $\langle-,-\rangle:V\otimes V\longrightarrow\bbr$ satisfying
\begin{align}
\langle v_0,\ell_n(v_1,v_2,\dots,v_n)\rangle\=\pm\,\langle v_n,\ell_n(v_0,v_1,\dots,v_{n-1})\rangle
\end{align}
for all $n\geq1$ and $v_i\in V$.
Thus cyclic $L_\infty$-algebras generalize quadratic differential graded Lie algebras.
In the dual differential graded commutative algebra picture $(V^*,Q)$, a cyclic structure translates to a graded symplectic two-form $\omega\in\Omega^2(V[1])$  which is $Q$-invariant. 

An $L_\infty$-morphism $\Psi:\Lambda_V\longrightarrow\Lambda_{V'}$ between two cyclic $L_\infty$-algebras $(V,L,\langle-,-\rangle)$ and $(V',L',\langle-,-\rangle')$ is \emph{cyclic} if it preserves the inner products:
\begin{align}
\begin{split}
\langle\psi_1(v),\psi_1(w)\rangle' & \=\langle v,w\rangle \ , \\[4pt]
\sum_{i=1}^{n-1}\, \langle\psi_i(v_1,\dots,v_i),\psi_{n-i}(v_{i+1},\dots,v_n)\rangle' & \=0 \quad , \quad n>2 \ .
\end{split}
\end{align}

\paragraph{$\mbf{L_\infty}$-structure of classical field theory.}
Consider a classical field theory with generalized gauge symmetries defined on a manifold $M$. The Batalin-Vilkovisky (BV) formalism constructs a differential graded algebra  $(C_\bullet^\infty(V[1]),Q_{\textrm{\tiny BV}})$  on the graded vector space $V$ of BV fields consisting of ghosts, physical fields and antifields. The action of the BV differential $Q_{\textrm{\tiny BV}}$ on $V[1]$ is a polynomial in the ghosts, fields and antifields and their derivatives, which is dual to a sum over brackets $\ell_n$ on $V$. The BV symplectic form $\omega_{\textrm{\tiny BV}}$, which is dual to the antibracket, is of degree $-1$ on $V$ and it induces a cyclic inner product of degree $-3$. 

The homogeneous subspace $V_0$ contains the gauge parameters, $V_1$ the fields, $V_2$ the field equations, and $V_3$ the corresponding Noether identities. The negatively graded homogeneous subspaces $V_{-k}$ for $k\geq1$ encode `higher gauge transformations' (corresponding to ghosts-for-ghosts, and so on) for reducible symmetries, and dually $V_{k+3}$ contain the corresponding `higher Noether identities'. The brackets $\ell_n$ of the $L_\infty$-structure on $V$ then encode all information about the gauge symmetries and dynamics of the classical field theory in the following way.

{Gauge transformations} of fields  $A\in V_1$  by gauge parameters $\lambda\in V_0$ are given by
\begin{align}
\delta_\lambda A\=\ell_1(\lambda)+\ell_2(\lambda,A) - \tfrac12\,\ell_3(\lambda,A,A)+\cdots \ ,
\end{align}
where the ellipses denote higher order brackets involving higher orders of the fields $A$.
If the gauge symmetries close off-shell, then the {closure of gauge algebra} is encoded by the field-dependent gauge transformations
\begin{align}
[\delta_{\lambda_1},\delta_{\lambda_2}]A\=\delta_{-C(\lambda_1,\lambda_2;A)}A \ ,
\end{align}
where
\begin{align}
C(\lambda_1,\lambda_2;A)\=\ell_2(\lambda_1,\lambda_2)+\ell_3(\lambda_1,\lambda_2,A) + \cdots \ .
\end{align}
This follows from the homotopy relations of the $L_\infty$-algebra.

{The field equations} $\calf_A\=0$ are given by
\begin{align}
\calf_A\=\ell_1(A)-\tfrac12\,\ell_2(A,A)-\tfrac1{3!}\,\ell_3(A,A,A)+\cdots \ .
\end{align}
From the homotopy relations it follows that they are covariant:
\begin{align}
\delta_\lambda\calf_A \= \ell_2(\lambda,\calf_A) + \ell_3(\lambda,\calf_A,A) +\cdots \ .
\end{align}
The moduli space of classical physical states is the quotient of the space of solutions to the field equations by the gauge transformations. {Quasi-isomorphic $L_\infty$-algebras lead to isomorphic moduli spaces, and so give physically equivalent field theories in this sense}. The corresponding {Noether identities} are
\begin{align}
\cali_\lambda\ :=\ \ell_1(\calf_A)+\ell_2(\calf_A,A)-\tfrac12\,\ell_3(\calf_A,A,A)+\cdots\ \equiv \ 0 \ , 
\end{align}
which are identically zero \emph{off-shell} as a consequence of the homotopy relations. 

The cyclic structure of the $L_\infty$-algebra induces an action
\begin{align}
S\=\tfrac12\,\langle A,\ell_1(A)\rangle-\tfrac1{3!}\,\langle A,\ell_2(A,A)\rangle-\tfrac1{4!}\,\langle A,\ell_3(A,A,A)\rangle +\cdots \ .
\end{align}
Cyclicity implies that the Euler-Lagrange equations of this action are precisely $\calf_A\=0$, and that gauge invariance of $S$ is equivalent to the Noether identities $\cali_\lambda\=0$.

\paragraph{Example: Chern-Simons theory.}
The prototypical example is provided by Chern-Simons gauge theory. Let $M$ be a three-dimensional oriented manifold, and let $\frg$ be a quadratic Lie algebra invariant pairing $\Tr_\frg$. The underlying {cochain complex} of the $L_\infty$-structure is de~Rham complex of $\frg$-valued differential forms $V\=\Omega^\bullet(M,\frg)$ with differential $\ell_1\=\dd$. The only other non-zero brackets are given by $\ell_2\=-[-,-]_\frg$, the Lie bracket of $\frg$ composed with exterior multiplication of differential forms. The cyclic structure of degree~$-3$ is defined by the inner product 
\begin{align}
\displaystyle{\langle\alpha,\beta\rangle\=\int_M\Tr_\frg(\alpha\wedge\beta)} \ ,
\end{align}
for forms $\alpha,\beta\in\Omega^\bullet(M,\frg)$ in complementary degrees.

The gauge transformation of a field $A\in V_1\=\Omega^1(M,\frg)$ by a gauge parameter $\lambda\in V_0\=\Omega^0(M,\frg)$ is then the standard one:
\begin{align}
\delta_\lambda A \= \dd\lambda - [\lambda,A]_\frg \ ,
\end{align}
which closes in the Lie algebra $(\Omega^0(M,\frg),[-,-]_\frg)$.
{The field equations} $\calf_A\=0$ are given by
\begin{align}
\calf_A\=F_A \ := \ \dd A +\tfrac12\,[A,A]_\frg \qquad \mbox{with} \quad \delta_\lambda F_A\=-[\lambda,F_A]_\frg \ ,
\end{align}
and the corresponding moduli space of classical physical states is the moduli space of flat connections on $M$. The {Noether identity} is simply the Bianchi identity
\begin{align}
\cali_\lambda\=\dd F_A-[F_A,A]_\frg\=0 \ .
\end{align}

The action reproduces the Chern-Simons functional
\begin{align}
\displaystyle{S\=\int_M \Tr_\frg\Big(\frac12\,A\wedge \dd A +\frac1{3!}\,A\wedge[A,A]_\frg \Big)} \ .
\end{align}
Thus {Chern-Simons gauge theory is organised by a differential graded Lie algebra}.

\section{Braided $L_\infty$-algebras of noncommutative field theories}
\label{sec:braided}

The standard noncommutative field theories with star-gauge symmetry have $L_\infty$-structures with finitely-many brackets which simply follow from the analogue structures of classical non-abelian gauge theories~\cite{Giotopoulos:2021ieg}. In this section we will instead define noncommutative field theories with {\em braided gauge symmetries}. We will discuss how to formulate them through a general notion of {\em braided ${L_\infty}$-algebras}, and illustrate how to use this to systematically construct new examples of noncommutative gauge theories.

\paragraph{Braided noncommutative deformation.}
Let $U\Gamma(TM)$ be the universal enveloping algebra of the Lie algebra $\Gamma(TM)$ of vector fields on a manifold $M$, endowed with its canonical cocommutative Hopf algebra structure.
Let  $\CCF\={\rm f}^\alpha\otimes {\rm f}_\alpha \in U\Gamma(TM)\otimes U\Gamma(TM)$  be a \emph{Drinfel'd twist}, with inverse $\CCF^{-1}\=\bar{\rm f}^\alpha\otimes \bar{\rm f}_\alpha$ and with $\CCF_{21}\={\rm f}_\alpha\otimes {\rm f}^\alpha$ the twist with its legs in $U\Gamma(TM)$ swapped. We do not review the theory of Drinfel'd twists here, and for concreteness the reader may wish to simply consider the standard {Moyal-Weyl twist} 
\begin{align}\label{eq:MWtwist}
\CCF\=\exp\big(-\tfrac\ii2\,\theta^{\mu\nu}\,\partial_\mu\otimes\partial_\nu\big) \ ,
\end{align}
which provides the simplest example of the general formalism (see e.g.~\cite{Giotopoulos:2021ieg} for a more detailed account with many further examples).

Physical fields are described by functions, forms and tensors on $M$, and so are elements of an algebra $\alg$ on which $\Gamma(TM)$ acts via the Lie derivative and the Leibniz rule; that is, $\alg$ is a $U\Gamma(TM)$-module algebra. Using the twist we deform the product $\,\cdot\,$ on $\alg$ into an associative star-product
\begin{align}
a\star b\=\,\boldsymbol\cdot \ \CCF^{-1}(a\otimes b) \= \bar {\rm f}^\alpha(a)\cdot\bar {\rm f}_\alpha (b) \ .
\end{align}
This defines a noncommutative algebra $\alg_\star$ carrying a representation of a twisted Hopf algebra $U_\CCF\Gamma(TM)$.
If $\alg$ is a commutative algebra, then $\alg_\star$ is \emph{braided}-commutative:
\begin{align}
a\star b\=\bar {\rm R}^\alpha(b)\star \bar {\rm R}_\alpha (a) \ ,
\end{align}
where $\CCR\=\CCF_{21}\,\CCF^{-1}\={\rm R}^\alpha\otimes {\rm R}_\alpha$ is the triangular $R$-matrix with inverse $\CCR^{-1}\=\bar{\rm R}^\alpha\otimes \bar{\rm R}_\alpha$. 

\paragraph{Braided gauge symmetry.}
Let $\frg$ be a Lie algebra with bracket $[-,-]_\frg$, and apply the twist deformation procedure to the Lie algebra $(\Omega^0(M,\frg),[-,-]_\frg)$ to get the new bracket
\begin{align}
[\lambda_1,\lambda_2]_\frg^\star\ :=\ [-,-]_\frg\circ\CCF^{-1}(\lambda_1\otimes \lambda_2) \= [\bar{\rm f}^\alpha(\lambda_1),\bar{\rm f}_\alpha(\lambda_2)]_\frg \ .
\end{align}
The resulting structure $\Omega_\star^0(M,\frg)$ is no longer a Lie algebra but rather a
\emph{braided Lie algebra}~\cite{Woronowicz:1989rt,Majid:1993yp}; that is, it is braided antisymmetric and obeys the braided Jacobi identity: 
\begin{align}\label{StarBracket2}
\begin{split}
[\lambda_1,\lambda_2]_\frg^\star & \=-[\bar {\rm R}^\alpha(\lambda_2),\bar {\rm R}_\alpha (\lambda_1)]_\star \ , \\[4pt][\lambda_1,[\lambda_2,\lambda_3]_\frg^\star]_\frg^\star & \=[[\lambda_1,\lambda_2]_\frg^\star,\lambda_3]_\frg^\star + [\bar{\rm R}^\alpha(\lambda_2),[\bar{\rm R}_\alpha(\lambda_1),\lambda_3]_\frg^\star]_\frg^\star \ .
\end{split}
\end{align}
Note that for a matrix Lie algebra $\frg$, the braided commutator is \emph{not} the same as the more conventional star-commutator:
\begin{align}
[\lambda_1,\lambda_2]_\frg^\star\=\lambda_1\star\lambda_2 - \bar{\rm R}^\alpha(\lambda_2)\star\bar{\rm R}_\alpha(\lambda_1) \ \neq \ \lambda_1\star\lambda_2 - \lambda_2\star\lambda_1 \= [\lambda_1\! \stackrel{\scriptstyle\star}{\scriptstyle,}\! \lambda_2]_{\frg} \ .
\end{align}

Let $W$ be a linear representation of the Lie algebra $\frg$. {Braided gauge fields $A\in\Omega_\star^1(M,\frg)$ and braided matter fields $\phi\in\Omega_\star^p(M,W)$  transform in (left) braided representations
\begin{align}
\delta_\lambda^{\star} \phi\=-\lambda\star \phi \qquad \mbox{and} \qquad \delta_\lambda^{\star}A\=\dd\lambda-[\lambda,A]_\frg^\star  \ .
\end{align}
These braided gauge transformations are not derivations but rather follow a \emph{braided Leibniz rule}, for instance
\begin{align}
\delta_\lambda^\star(\phi\otimes A)\=\delta_\lambda^\star \phi\otimes A+\bar {\rm R}^\alpha (\phi)\otimes\delta^\star_{\bar {\rm R}_\alpha (\lambda)}A \ ,
\end{align}
and they close a braided Lie algebra defined by the braided commutator
\begin{align}
\big[\delta_{\lambda_1}^\star,\delta_{\lambda_2}^\star\big]^\star \ := \ \delta_{\lambda_1}^\star\,\delta_{\lambda_2}^\star  - \delta_{\bar{\rm R}^\alpha(\lambda_2)}^\star\,\delta^\star_{\bar{\rm R}_\alpha(\lambda_1)} \= \delta^\star_{[\lambda_1,\lambda_2]_\frg^\star}  \ .
\end{align}

Braided gauge invariant dynamics is built from the braided left and right covariant derivatives
\begin{align}
\dd_{\star\lact}^A\phi \ := \ \dd \phi+A\wedge_\star \phi \qquad \mbox{and} \qquad \dd_{\star\ract}^A\phi \ := \ \dd \phi+\bar{\rm R}^\alpha(A)\wedge_\star \bar{\rm R}_\alpha(\phi) \ ,
\end{align}
which are braided covariant: 
\begin{align}
\delta_\lambda^\star\big( \dd_{\star\lact,\ract}^A\phi\big) \= -\lambda\star\big( \dd_{\star\lact,\ract}^A\phi\big) \ . 
\end{align}
In particular, the {braided curvature} of a gauge field is
\begin{align}
F_A^\star \ := \ \dd A+\tfrac12\,[A,A]_\frg^\star \qquad \mbox{with} \quad \delta_\lambda^\star F_A^\star\=-[\lambda,F_A^\star]_\frg^\star\ .
\end{align}

In our applications to gravity later on, we are particularly interested in the deformation of the Lie algebra of vector fields $(\Gamma(TM),[-,-])$, which defines the braided Lie algebra of braided diffeomorphisms $\Gamma_\star(TM)$. We construct a noncommutative differential geometry on $M$ by requiring it to be covariant with respect to the twisted Hopf algebra $U_\CCF\Gamma(TM)$~\cite{Aschieri:2005zs,SpringerBook,Aschieri:2017sug}. The action of a twisted diffeomorphism, generated by a vector field $\xi\in\Gamma(TM)$, on a tensor field $\mbf T$ is given by the braided Lie derivative
\begin{align}\label{StarLieDer}
\CL_\xi^\star \mbf T \ := \ \CL_{\bar {\rm f}^\alpha(\xi)}\bar {\rm f}_\alpha  \mbf(\mbf T) \qquad \mbox{with} \quad \big[\CL_{\xi_1}^\star,\CL_{\xi_2}^\star\big]^\star \= \CL_{[\xi_1,\xi_2]^\star}^\star \ .
\end{align}

\paragraph{Braided $\mbf{L_\infty}$-structure of braided field theory.}
Let $(V,\{\ell_n\})$ be a classical $L_\infty$-algebra in the category of $U\Gamma(TM)$-modules; this means that $U\Gamma(TM)$ acts on the graded vector space $V$ and all brackets $\ell_n$ are $U\Gamma(TM)$-equivariant maps. Via the twist deformation procedure we obtain new brackets $\ell_n^\star$ defined by
\begin{align}
\ell_n^\star(v_1\wedge\cdots\wedge v_n) \ := \ \ell_n(v_1\wedge_\star\cdots\wedge_\star v_n) \ .
\end{align}
Note that the differential $\ell_1^\star\=\ell_1$ is unchanged, and hence so is the underlying cochain complex $(V,\ell_1)$; this will translate to the statement that the free classical field theory in unchanged by the noncommutative deformation. The maps $\ell_n^\star$ for $n\geq2$ are \emph{braided graded antisymmetric:}
\begin{align}
\ell_n^\star(\dots,v,v',\dots)\=-(-1)^{|v|\,|v'|} \ \ell_n^\star(\dots,\bar {\rm R}^\alpha(v'),\bar {\rm R}_\alpha (v),\dots) \ ,
\end{align}
and they satisfy \emph{braided homotopy Jacobi identities}, which are unchanged for $n\=1,2$. In~\cite{Ciric:2020eab,Ciric:2021rhi,Giotopoulos:2021ieg} the resulting algebraic structure $(V,\{\ell_n^\star\})$ is called a \emph{braided $L_\infty$-algebra} in the category of $U_\CCF\Gamma(TM)$-modules. 

The braided $L_\infty$-structure can be used to build noncommutative field theories with braided symmetries in the same way as in the classical case. The {braided gauge transformations}
\begin{align}
\delta_\lambda^\star A\=\ell^\star_1(\lambda) + \ell_2^\star(\lambda,A)-\tfrac12\,\ell_3^\star(\lambda,A,A)+\cdots
\end{align}
close a braided Lie algebra under the braided commutator ${[-,-]^\star}$. {The braided field equations} 
\begin{align}
\calf_A^\star\=\ell_1^\star(A) - \tfrac12\,\ell_2^\star(A,A) -\tfrac1{3!}\,\ell_3^\star(A,A,A) + \cdots
\end{align}
are covariant:
\begin{align}
\delta_\lambda^\star \calf_A^\star\=\ell_2^\star(\lambda,\calf_A^\star) + \tfrac12\,\big(\ell_3^\star(\lambda,\calf_A^\star,A) - \ell_3^\star(\lambda,A,\calf_A^\star)\big) + \cdots \ .
\end{align}
Although there is no conventional moduli space of solutions to $\calf_A^\star\=0$~\cite{Ciric:2021rhi,Giotopoulos:2021ieg}, a set of \emph{braided Noether identities} continue to hold from a weighted sum over all braided homotopy identities evaluated on $(A^{ n})$:
\begin{align}
\begin{split}
{\cali_A^\star \calf_A^\star} & \=  {\ell_1^\star(\calf_A^\star) + \tfrac12\,\big(\ell_2^\star(\calf_A^\star,A)-\ell_2^\star(A,\calf_A^\star)\big)} \\ & \qquad  {+ \tfrac1{3!} \, \ell_1^\star\big(\ell_3^\star(A,A,A)\big) + \tfrac14\,\big(\ell_2^\star(\ell_2^\star(A,A),A)-\ell_2^\star(A,\ell_2^\star(A,A))\big) + \cdots \ \equiv \ 0} \ .
\end{split}
\end{align}

If the classical $L_\infty$-algebra $(V,\{\ell_n\})$ is further equipped with a $U\Gamma(TM)$-invariant cyclic inner product $\langle-,-\rangle:V\otimes V\longrightarrow\bbr$ of degree~$-3$, then
\begin{align}
\langle-,-\rangle_\star \ := \ \langle-,-\rangle\circ \CCF^{-1} \= \langle\bar{\rm f}^\alpha(-),\bar{\rm f}_\alpha(-)\rangle
\end{align}
is a cyclic inner product on the braided $L_\infty$-algebra $(V,\{\ell^\star_n\})$ of degree~$-3$. With it we construct the action
\begin{align}
S^\star\=\tfrac12\,\langle A,\ell_1^\star(A)\rangle_\star - \tfrac1{3!}\,\langle A,\ell_2^\star(A,A)\rangle_\star -\tfrac1{4!}\,\langle A,\ell_3^\star(A,A,A)\rangle_\star + \cdots \ .
\end{align}
As in the classical case, cyclicity implies that $\delta S^\star\=\langle\delta A,\calf_A^\star\rangle_\star$ for arbitrary field variations $\delta A$, and that $\delta_\lambda^\star S^\star\=-\langle\lambda,\cali_A^\star \calf_A^\star\rangle_\star$ for arbitrary braided gauge transformations. Hence the variational principle for the action $S^\star$ leads to the braided field equations $\calf_A^\star\=0$, while braided gauge invariance of $S^\star$ is equivalent to the braided Noether identities $\cali_A^\star \calf_A^\star\=0$. For the relation of this construction to a braided version of the BV formalism, see~\cite{Nguyen:2021rsa,Giotopoulos:2021ieg}.

The formulation provided by braided $L_\infty$-algebras yields systematic constructions of \emph{new} noncommutative field theories with no new degrees of freedom and good classical limits, unlike other approaches to noncommutative field theories with arbitrary gauge algebras and matter field representations (see e.g.~\cite{Giotopoulos:2021ieg} for a recent review), as well as some ``surprises''. Let us briefly mention some standard field theory examples; for further details and examples, see~\cite{Giotopoulos:2021ieg}. Aspects of the corresponding braided quantum field theories are currently under investigation.

\paragraph{Example I: Braided Chern-Simons theory.} 
The prototypical example is braided Chern-Simons theory on a three-dimensional oriented manifold $M$ with quadratic Lie algebra $(\frg,[-,-]_\frg,\Tr_\frg)$ which is built on a differential graded braided Lie algebra with homogeneous subspaces $V_p\=\Omega^p(M,\frg)$, $p=0,1,2,3$ and the non-vanishing brackets
\begin{align}
\ell_1^\star\=\ell_1\=\dd \qquad \mbox{and} \qquad \ell_2^\star\=-[-,-]_\frg^\star \ .
\end{align}
The action
\begin{align}
S^\star\=\int_M\Tr_\frg\Big(\frac12\,A\wedge_\star \dd A +\frac1{3!}\,A\wedge_\star[A,A]^\star_\frg \Big)
\end{align}
{is invariant} under braided gauge transformations closing the braided Lie algebra $\Omega^0_\star(M,\frg)$. {The field equations} are $F_A^\star\=0$ and they are braided covariant. The
{Bianchi identities} are modified, see~\cite{Ciric:2020eab,Giotopoulos:2021ieg}.

\paragraph{Example II: Noncommutative scalar field theory.}
The $L_\infty$-algebra formalism is also equipped to handle theories without gauge symmetries~\cite{BVChristian,Giotopoulos:2021ieg}. Consider, for example, massive scalar $\kappa\,\phi^3$-theory on $d$-dimensional Minkowski space $\bbr^{1,d-1}$. In this case $V_0\=V_3\=0$ since there are no gauge symmetries, while $V_1\=V_2\= C^\infty(\bbr^{1,d-1})$ with the non-vanishing brackets
\begin{align}
\ell_1^\star\=\ell_1\= \square-m^2 \qquad \mbox{and} \qquad \ell^\star_2(\phi_1,\phi_2)\= \kappa\,\phi_1\star\phi_2
\end{align}
for $\phi_1,\phi_2\in V_1$. Note that $\ell_2^\star$ is braided symmetric, and that the braided $L_\infty$-relations hold trivially for degree reasons. The {braided field equations are}
\begin{align}
\calf_\phi^\star\=\ell_1^\star(\phi)-\tfrac12\,\ell_2^\star(\phi,\phi)\=(\square-m^2)\,\phi-\tfrac\kappa2\,\phi\star\phi \ .
\end{align}

With the cyclic inner product 
\begin{align}
\langle\phi,\phi^+\rangle_\star\=\int\dd^{d} x \ \phi \star\phi^+
\end{align}
for $\phi\in V_1$ and $\phi^+\in V_2$, the {action} is
\begin{align}
S^\star\=\int\dd^{d} x \ \frac12\,\phi\star(\square-m^2)\,\phi-\frac\kappa{3!}\,\phi\star\phi\star\phi \ .
\end{align}
It follows that {\it standard} noncommutative scalar field theory is organised by a {\it braided} $L_\infty$-algebra. Like the other usual noncommutative field theories~\cite{Giotopoulos:2021ieg}, it also has a standard $L_\infty$-algebra formulation in terms of the symmetric bracket
\begin{align}
\ell_2^{\star{\rm sym}}(\phi_1,\phi_2) \= \tfrac\kappa2 \, (\phi_1\star\phi_2 + \phi_2\star\phi_1) \ .
\end{align}
The differences between the two homotopy algebraic formulations appear in the respective quantum field theories, see~\cite{Nguyen:2021rsa}.

\paragraph{Example III: Braided noncommutative Yang-Mills theory.}
Starting from the standard Yang-Mills action 
\begin{align}
{S\=\frac12\,\int\Tr_\frg(F_A\wedge *\,F_A)}
\end{align}
for a gauge field $A\in\Omega^1(\bbr^{1,d-1},\frg)$, the classical Yang-Mills $L_\infty$-algebra is not a differential graded Lie algebra, because $\ell_3(A,A,A) \ \neq \ 0$~\cite{BVChristian,Giotopoulos:2021ieg}, unlike the previous two examples. The corresponding braided Yang-Mills $L_\infty$-algebra gives the {braided field equations}
\begin{align}
\begin{split}
& \tfrac{1}{2}\, \big(\dd^A_{\star\lact}\ast  F_A^{\star} + \dd^A_{\star\ract}\ast  F_A^{\star}\big) \\ & \quad +\, \tfrac{1}{6} \, \big[\bar{\rm R}^\alpha(A),\, \ast\, [\bar{\rm R}_\alpha (A),A]_\frg^\star \,\big]_\frg^\star
		-\tfrac{1}{12} \,  \big[A,\, \ast\, [A,A]_\frg^\star\, \big]_\frg^\star + \tfrac{(-1)^{d}}{12}\, \big[ \ast [A,A]_\frg^\star\, ,A\big]_\frg^\star \= 0 \ .
\end{split}
\end{align}
Despite the appearence of the second line, these are covariant and the classical limit gives the usual Yang-Mills equations  $\dd^A\ast F_A\=0$~\cite{Giotopoulos:2021ieg}. The action of the braided gauge theory is
\begin{align}
\begin{split}
S^\star & \=  \frac{1}{2}\, \int \Tr_\frg \big(F_A^\star \wedge_\star  \ast\, F_A^\star\big) \\
& \qquad
+ \frac{1}{24}\, \int \Tr_\frg \big([A,\bar{\rm R}^\alpha(A)]_\frg^\star \, \wedge_\star \, \ast\, [\bar{\rm R}_\alpha(A),A]_\frg^\star\,  \big) -[A,A]_\frg^\star \, \wedge_\star \, \ast\,  [A,A]_\frg^\star\,  \big) \ .
\end{split}
\end{align}
Again this is gauge invariant with a good classical limit. The braided Noether identities are a complicated modification of the classical identities $\dd^A(\dd^A\ast F_A)\=0$. This noncommutative field theory defines a new deformation of Yang-Mills theory. For further details and discussion, see~\cite{Giotopoulos:2021ieg}. 

\section{$L_\infty$-algebras of general relativity in the first order formalism}
\label{sec:GR}

In the remainder of this contribution we turn our attention to the applications of the formalism from Sections~\ref{sec:LinftyCFT} and~\ref{sec:braided} to gravity. The $L_\infty$-structure of classical gravity without matter fields in the Einstein-Cartan-Palatini (ECP) formulation was discussed in detail in~\cite{Ciric:2020hfj}, in arbitrary dimensionality and spacetime signature. In this section we briefly review this construction, focusing for concreteness on the gravity theories in three and four dimensions as well as with Lorentzian signature.

\paragraph{Einstein-Cartan-Palatini gravity.}
Let $M$ be a $d$-dimensional oriented manifold. Let $\CV\longrightarrow M$ be an oriented vector bundle of rank $d$ which is isomorphic to the tangent bundle $TM$ and is endowed with a fibrewise Minkowski metric $\eta$; the bundle $\CV$ is called the `fake tangent bundle'. The field content of ECP theory on $M$ consists of a bundle isomorphism $e:TM\longrightarrow \CV$, which defines the coframe field $e\in\Omega^1(M,\CV)$, and a connection on the associated principal $SO(1,d-1)$-bundle $P\longrightarrow M$ to $\CV$ called the spin connection, which has a curvature two-form 
\begin{align}
R\=\dd\omega+\tfrac12\,[\omega,\omega]\in \Omega^2(M,P\times_{\rm ad}\mathfrak{so}(1,d-1))\ . 
\end{align}

These fields satisfy the {Bianchi identities}  
\begin{align}
\dd^\omega T\=R\wedge e \qquad \mbox{and} \qquad \dd^\omega R\=0 \ ,
\end{align}
where
\begin{align}
T\=\dd^\omega e\=\dd e+\omega\wedge e
\end{align}
is the torsion of $\omega$. The orientation of $\CV$ induces a map $\Tr:\Omega^d(M,\wedge^d\CV)\longrightarrow \Omega^d(M)$. Locally, or globally if $M$ is parallelizable, we take $e \in \Omega^1(M,\bbr^{1,d-1})$, $\omega \in \Omega^1(M,\mathfrak{so}(1,d-1))$ with $\aso(1,d-1) \ \simeq \ \wedge^2(\bbr^{1,d-1})$, and $\Tr:\wedge^d(\bbr^{1,d-1})\longrightarrow\bbr$. We shall always work in this local situation.

In any dimension $d$, the action for ECP gravity with cosmological constant $\Lambda\in\bbr$ is given by
\begin{align}
\begin{split}
S & \=\int_M \Tr\Big(\frac1{d-2}\,e^{d-2}\dwedge R + \frac{\Lambda}d\,e^{d}\Big) \\[4pt]
& \=\int_{M}\, \varepsilon_{a_1\cdots a_d}\, \Big(e^{a_1}\wedge\cdots\wedge e^{a_{d-2}}\wedge R^{a_{d-1}a_d} + \frac\Lambda d \, e^{a_1}\wedge\cdots\wedge e^{a_d}\Big) \ ,
\end{split}
\end{align}
where the $\dwedge$-product is the double exterior product combining the differential form exterior product in $\Omega^\bullet(M)$ with the multivector exterior product in $\wedge^\bullet(\bbr^{1,d-1})$, while the $\wedge$-product acts via multivector representations and exterior multiplication of spacetime forms.
The infinitesimal gauge symmetries of this theory are given by the semi-direct product Lie algebra of diffeomorphisms and local Lorentz transformations:
\begin{align}\label{eq:classsemidirect}
\Gamma(TM)\ltimes\Omega^0(M,\mathfrak{so}(1,d-1)) \ .
\end{align}

{The corresponding field equations are}
\begin{align}
e^{d-3}\dwedge T\=0 \qquad \mbox{and} \qquad e^{d-3}\dwedge R + \Lambda\, e^{d-1} \=0 \ .
\end{align}
In three and four dimensions, these equations respectively read
\begin{align}
\begin{split}
T\=0\=R + \Lambda\,e\dwedge e \ \ \ \mbox{\sf (d\,=\,3)} \qquad \mbox{,} \qquad e\dwedge T\=0\=e\dwedge R+\Lambda\, e\dwedge e\dwedge e \ \ \ \mbox{\sf (d\,=\,4)} \ .
\end{split}
\end{align}
For non-degenerate coframes $e$, these are respectively equivalent to the torsion-free condition for the spin connection and the vacuum Einstein equations with cosmological constant.

Note that in contrast to the (second order) Einstein-Hilbert formulation of general relativity, the (first order) ECP theory makes sense for degenerate coframes $e$, as required for its $L_\infty$-algebra formulation which is based on linear spaces. In the following we present this $L_\infty$-structure for the case of parallelizable spacetimes in three and four dimensions. See~\cite{Ciric:2020hfj} for a more general construction of `covariant $L_{\infty}$-algebras' for arbitrary spacetimes, which are quasi-isomorphic to the $L_\infty$-algebras below in the case of parallelizable manifolds.

\paragraph{$\mbf{L_\infty}$-structure in $\mbf{d=3}$.}
{The cochain complex underlying three-dimensional gravity} is
\begin{align}
V_0\xrightarrow{ \ \ \ell_1 \ \ }V_1\xrightarrow{ \ \ \ell_1 \ \ 
  }V_2\xrightarrow{ \ \ \ell_1 \ \ }V_3 \ ,
\end{align}
where the various vector spaces encode the symmetries and dynamics as follows:
\begin{align}
\begin{split}
\mbox{\sf Gauge transformations:}& \hspace{2cm} {(\xi,\lambda)}\in \  V_0\=\Gamma(TM)\times\Omega^0(M,\aso(1,2)) \\[4pt]
\mbox{\sf Physical fields:}& \hspace{2cm}  {(e,\omega)}\in   \ V_1\=\Omega^1(M,\bbr^{1,2})\times\Omega^1(M,\aso(1,2)) \\[4pt]
\mbox{\sf Field equations:}& \hspace{1.9cm} {(E,\varOmega)}\in  \ V_2\= \Omega^2(M,\wedge^2(\bbr^{1,2}))\times\Omega^2(M,\bbr^{1,2}) \\[4pt]
\mbox{\sf Noether identities:}& \hspace{2cm}  {(\varXi,\varLambda)}\in  \ V_3\=\Omega^1(M,\Omega^3(M))\times\Omega^3(M,\bbr^{1,2})
\end{split}
\end{align}

{The differential is given by} 
\begin{align}
\ell_1(\xi,\lambda)\=(0,\dd\lambda) \  , \quad  \ell_1(e,\omega)\=(\dd\omega,\dd e) \qquad \mbox{and} \qquad \ell_1(E,\varOmega)\=(0,\dd\varOmega) \ .
\end{align}
{The first few $2$-brackets are}  
\begin{align}\label{eq:d=3ell2}
\begin{split}
\ell_2((\xi_1,\lambda_1),(\xi_2,\lambda_2))&\=([\xi_1,\xi_2],-[\lambda_1,\lambda_2]+\CL_{\xi_1}\lambda_2-\CL_{\xi_2}\lambda_1) \ , \\[4pt]
\ell_2((\xi,\lambda),(e,\omega))&\=(-\lambda\cdot e+\CL_\xi e,-[\lambda,\omega]+\CL_\xi\omega) \ , \\[4pt]
\ell_2((e_1,\omega_1),(e_2,\omega_2))&\=-([\omega_1,\omega_2] + 2\,\Lambda\,e_1\dwedge e_2,\omega_1\wedge e_2+\omega_2\wedge e_1) \ ,
\end{split}
\end{align}
along with three further $2$-brackets related to the field equations and Noether identities. There are no higher brackets, and so three-dimensional gravity is organised by a differential graded Lie algebra.

{Explicitly, the gauge symmetry is encoded by}
\begin{align}
\delta_{(\xi,\lambda)}(e,\omega)\=(-\lambda\cdot e+\CL_\xi e\,,\,\dd\lambda-[\lambda,\omega]+\CL_\xi\omega)\=\ell_1(\xi,\lambda)+\ell_2((\xi,\lambda),(e,\omega)) \ ,
\end{align}
{and the field equations by} 
\begin{align}
\calf_{(e,\omega)}\=(\calf_e,\calf_\omega)\=(R+\Lambda\,e\dwedge e,T)\=\ell_1(e,\omega)-\tfrac12\,\ell_2((e,\omega),(e,\omega)) \ .
\end{align}
The Noether identities are
\begin{align}
\begin{split}
\cali_{(\xi,\lambda)}&\=\big(\dd x^\mu\otimes\Tr(\iota_{\partial_\mu}e\dwedge\dd \calf_\omega-\iota_{\partial_\mu}\dd e\dwedge \calf_\omega)+(e\leftrightarrow\omega)\,,\,\dd^\omega T-R\wedge e\big)\\[4pt] &\=\ell_1(\calf_{(e,\omega)})-\ell_2((e,\omega),\calf_{(e,\omega)}) \ \equiv \ (0,0)  \ ,
\end{split}
\end{align}
and they are identically equal to zero \emph{off-shell}.

{The cyclic inner product}  
\begin{align}\label{ActionII}
\langle(e,\omega),(E,\varOmega)\rangle \ := \ \int_M\Tr(e\dwedge E+\varOmega\dwedge\omega)
\end{align}
on $V_1\otimes V_2$ encodes the ECP action
\begin{align}
S\=\tfrac12\,\langle(e,\omega)\,,\,\ell_1(e,\omega)-\tfrac1{3!}\,\ell_2((e,\omega),(e,\omega))\rangle \ .
\end{align}
We can extend this pairing to  $\langle-,-\rangle:V_0\otimes V_3\longrightarrow\bbr$  using gauge invariance and integration by parts. Then cyclicity on $V_0$ implies the Noether identities.

\paragraph{Application to Chern-Simons gravity.}
Three-dimensional gravity is equivalent to a Chern-Simons theory whose gauge algebra $\frg$ is the Poincar\'e algebra $\mathfrak{iso}(1,2)\=\bbr^{1,2}\rtimes\aso(1,2)$ for $\Lambda\=0$, the de~Sitter algebra $\aso(1,3)$ for $\Lambda>0$, or the anti-de~Sitter algebra $\aso(2,2)$ for $\Lambda<0$~\cite{Witten:1988hc}. The field content of ECP theory may be assembled into a connection one-form $A\=(e,\omega)\in\Omega^1(M,\frg)$ with curvature two-form  $F_A\=(R,T)\in\Omega^2(M,\frg)$ in this Chern-Simons formulation.

A key point of this correspondence is the manner in which the diffeomorphism symmetry of general relativity is realised on the Chern-Simons side. For invertible coframes $e$, diffeomorphisms generated by vector fields $\xi\in\Gamma(TM)$  are equivalent \emph{on-shell} to field-dependent gauge transformations by 
\begin{align}
(\tau_\xi,\lambda_\xi)\=(\iota_\xi e,\iota_\xi\omega)\in\Omega^0(M,\frg) \ .
\end{align}
Explicitly,
\begin{align}
\delta_\xi A\=\CL_\xi A\=\delta_{(\tau_\xi,\lambda_\xi)}A + \iota_\xi F_A \ .
\end{align}
However, if one allows for degenerate metrics, as is necessary in the $L_\infty$-algebra formalism, then the gauge orbits under the ``shifts'' $(\tau_\xi,\lambda_\xi)$ only strictly contain the diffeomorphism orbits on-shell. It was shown by~\cite{Ciric:2020hfj} how to extend the correspondence of~\cite{Witten:1988hc} to an \emph{off-shell} equivalence including arbitrary coframe fields.

Diffeomorphisms in Chern-Simons theory are \emph{redundant symmetries}. We can trivialise them via an extension $V_{\textrm{\tiny CS}}^{\rm ext}$ of the graded vector space $V_{\textrm{\tiny CS}}$ of Chern-Simons theory which augments the degree~$0$ (resp. degree~$3$) subspace by $\Gamma(TM)$ (resp. $\Omega^1(M,\Omega^3(M))$). On the ECP side  we extend the homogenous subspace $V_0$ (resp. $V_3$) by the extra ``shift'' symmetries $\Omega^0(M,\bbr^{1,2})$ (resp.~$\Omega^3(M,\aso(1,2))$) to get  a new graded vector space $V^{\rm ext}_{\textrm{\tiny ECP}}$. In both cases we then add the subspace $V_{-1}\=\Gamma(TM)$ (resp. $V_4\=\Omega^1(M,\Omega^3(M))$  with  differential $\ell_1:V_{-1}\hookrightarrow V_0$ (resp.~$\ell_1:V_3\twoheadrightarrow V_4$). The underlying cochain complexes $(V_{\textrm{\tiny CS}},\ell_1)$ and $(V_{\textrm{\tiny CS}}^{\rm ext},\ell_1)$ then have the same cohomology, and altogether we get an isomorphism
\begin{align}
H^\bullet(V_{\textrm{\tiny ECP}}^{\rm ext},\ell_1) \ \simeq \ H^\bullet(V_{\textrm{\tiny CS}},\ell_1) \ .
\end{align}
In~\cite{Ciric:2020hfj} it is shown that there is an (off-shell) cyclic $L_\infty$-quasi-isomorphism $\{\psi_n\}_{n\geq1}$  with $\psi_n\=0$ for $n\geq3$ from the differential graded Lie algebra of Chern-Simons theory to the differential graded Lie algebra of ECP theory.

\paragraph{$\mbf{L_\infty}$-structure in $\mbf{d=4}$.}
The  graded vector space underlying four-dimensional gravity is now given by
\begin{align} 
\begin{split}
V_{0}&\=\Gamma(TM)\times \Omega^{0}\big(M,\mathfrak{so}(1,3)\big) \ , \\[4pt] 
V_{1}&\= \Omega^{1}(M,\bbr^{1,3}) \times
\Omega^{1}(M,\mathfrak{so}(1,3) ) \ , \\[4pt]
V_{2}&\=\Omega^{3}(M,\wedge^{3}(\bbr^{1,3})) \times
\Omega^{3}(M,\wedge^{2}(\bbr^{1,3})) \ , \\[4pt]
V_3&\=\Omega^1(M,\Omega^4(M)) \times \Omega^4(M,\wedge^2(\bbr^{1,3}))
\ .
\end{split}
\end{align} 
The differential is
\begin{align} \label{eq:ell14d}
\ell_{1}(\xi,\lambda)\=(0,\dd\lambda) \ , \quad
\ell_{1}(e,\omega)\=(0,0) \  \qquad \mbox{and} \qquad \ell_{1}(E,{\varOmega})\=(0,-\dd {\varOmega}) \ .
\end{align}
The only  $2$-bracket which is modified from the list \eqref{eq:d=3ell2} is
\begin{align}
\ell_{2}((e_{1},\omega_{1}),(e_{2},\omega_{2}))\=-(e_{1} \dwedge \dd \omega_{2}
+ e_{2} \dwedge \dd \omega_{1} , e_{1}
\dwedge \dd e_{2} + e_{2} \dwedge \dd e_{1}) \ .
\end{align}
Finally, there is a single non-trivial $3$-bracket
\begin{align}
\begin{split}
& \ell_{3}((e_{1},\omega_{1}),(e_{2},\omega_{2}),(e_{3},\omega_{3})) \\[4pt]
& \hspace{1cm} \=-( e_{1}\dwedge [\omega_{2} , \omega_{3}] + e_{2} \dwedge [\omega_{1} , \omega_{3}] + e_{3} \dwedge [\omega_{2} , \omega_{1}] + 6\,\Lambda\,e_1\dwedge e_2\dwedge e_3, \\
& \hspace{2cm} \ e_{1} \dwedge (\omega_{2} \wedge e_{3}) 
+_{(2\leftrightarrow 3)} + e_{2}\dwedge (\omega_{1} \wedge e_{3}) +_{(1\leftrightarrow 3)} + e_{3}\dwedge (\omega_{2} \wedge e_{1})
+_{(2\leftrightarrow 1)}  ) \ .
\end{split}
\end{align}

Thus four-dimensional ECP gravity is organised by an {$L_\infty$-algebra} which is not a differential graded Lie algebra, as $\ell_3 \ \neq \ 0$, similarly to Yang-Mills theory. In~\cite{Ciric:2020hfj} it is shown that this $L_\infty$-algebra picture is dual to the BV--BRST formulation of ECP gravity from~\cite{Cattaneo:2017kkv}.

\section{$L_\infty$-algebras of noncommutative gravity}
\label{sec:LinftyNCgrav}

In this final section we consider noncommutative deformations of the ECP theory of gravity from Section~\ref{sec:GR}. After describing the issues surrounding its formulation in terms of the standard $L_\infty$-algebra picture of gauge symmetries and dynamics, we describe the braided $L_\infty$-algebra picture which leads to a new noncommutative deformation of gravity in four dimensions.

\paragraph{$\mbf{L_\infty}$-structure.}
Let us start by reviewing the standard formulation of noncommutative Einstein-Cartan-Palatini gravity, in order to elucidate the problems with formulating it in the standard picture of $L_\infty$-algebras. While the general features hold more generally, we shall focus here on the case of $d\=3$ spacetime dimensions without cosmological constant; this theory is discussed in~\cite{Aschieri:2009mc}. We will proceed in two steps: first we consider the noncommutative generalization of the local Lorentz transformations, and subsequently that of the diffeomorphism symmetry. We will then attempt to understand them together within the setting of classical $L_\infty$-algebras along the same lines as in Section~\ref{sec:GR}.

In the standard noncommutative gauge theories based on star-gauge symmetries, one can deform any theory which is based on a matrix gauge algebra that closes under both commutators and anticommutators (such as $\fru(N)$ and $\mathfrak{gl}(N,\bbc)$), and with matter fields in the trivial, (anti)fundamental or adjoint representations. These noncommutative field theories are organised by classical $L_\infty$-algebras in an identical way to those of classical non-abelian gauge theories, see~\cite{Giotopoulos:2021ieg}. For the local Lorentz symmetry, this would amount to defining the noncommutative transformation laws of the noncommutative coframe $\hat e$ and spin connection $\hat\omega$ by a gauge parameter $\hat\lambda$ as
\begin{align}\label{NCLorTransf}
\hat\delta_{\hat\lambda}\hat e \= -\hat\lambda\star\hat e \qquad \mbox{and} \qquad \hat\delta_{\hat\lambda}\hat \omega \= \dd\hat\lambda - [\hat\lambda \ds \hat\omega] \= \dd\hat\lambda - (\hat\lambda\star\hat\omega - \hat\omega\star\hat \lambda) \ ,
\end{align}
where the noncommutative fields and gauge parameters are designated with a hat.

However, the Lorentz algebra $\mathfrak{so}(1,2)$ is not closed under the star-commutator $[- \ds -]$. Generally, for an arbitrary matrix Lie algebra $\frg$ one solves this problem by enlarging $\frg$ to its universal enveloping algebra $U\frg$, in which the star-commutator does close. For the special case $\frg\=\mathfrak{so}(1,2)$, the situation is simpler, because the anticommutator of two elements is always proportional to the identity matrix and so is a central element of $U\frg$: $\{\frg,\frg\}\subset\bbr\,\mbf{1}$. Hence it suffices to enlarge $\mathfrak{so}(1,2)$ to a centrally extended Lie algebra
\begin{align}\label{eq:centalext}
\hat\frg \= \mathfrak{so}(1,2) \oplus \bbr \ ,
\end{align}
whose fundamental representation is
\begin{align}\label{eq:centralextrep}
\hat W \= \bbr^{1,2} \oplus \bbr \ ,
\end{align}
and take all noncommutative fields and gauge parameters to be valued in $\hat\frg$ and $\hat W$. We denote these by $\hat\lambda\=\lambda+\tilde\lambda$, where $\lambda\in\Omega^0(M,\mathfrak{so}(1,2))$ and $\tilde\lambda\in\Omega^0(M)$, and similarly $\hat e\=e+\tilde e$ and $\hat \omega\=\omega+\tilde\omega$; we designate the new centrally extended variables with a tilde, while the classical variables from Section~\ref{sec:GR} remain unlabelled. 

This approach, which operates in the central extension \eqref{eq:centalext}--\eqref{eq:centralextrep}, introduces new fields $\tilde{e}$ and $\tilde{\omega}$. They can be regarded as new degrees of freedom which are not present in the classical theory. Another way to understand them is through the Seiberg-Witten map~\cite{Seiberg:1999vs}, which enables one to express all noncommutative fields as functionals of the corresponding commutative fields~\cite{Seiberg:1999vs,Jurco:2000ja}; this was applied to noncommutative ECP gravity in~\cite{Aschieri:2011ng,Aschieri:2012in}.

One can now check that the centrally extended local Lorentz transformations (\ref{NCLorTransf}) close the gauge algebra
\begin{equation}
\big[\hat\delta_{\hat{\lambda}_1}, \hat\delta_{\hat{\lambda}_2}\big] \= 
\hat\delta_{[{\hat{\lambda}}_1 \ds {\hat{\lambda}}_2]} \ . \label{AlgebraGaugeTransf}
\end{equation}
The noncommutative torsion and curvature are 
\begin{align}
\begin{split}
\hat{T}  & \= \dd \hat{e} + \tfrac12\,\big(\hat{\omega} \wedge_\star {\hat e} + \bar{\rm R}^\alpha(\hat\omega)\wedge_\star\bar{\rm R}_\alpha(\hat e)\big) \= T + \tilde{T} \ , \\[4pt]
\hat{R}  & \= \dd \hat{\omega} + \tfrac12\,[\hat{\omega}\ds \hat{\omega}] \= R + \tilde{R} \ .
\end{split}
\end{align}
In addition to (\ref{NCLorTransf}), one can show that they
transform covariantly under the infinitesimal noncommutative local Lorentz transformations:
\begin{eqnarray}
\hat\delta_{\hat{\lambda}} \hat{T} \= -\hat{\lambda} \star \hat{T} \qquad \mbox{and} \qquad
\hat\delta_{\hat{\lambda}} \hat{R} \= -[\hat{\lambda} \ds \hat{R}] \ . \label{NCTransfTR}
\end{eqnarray}

Calculating the star-operations explicitly, the components of $\hat T\in\Omega^2(M,\hat W)$ valued in $\bbr^{1,2}$ and $\bbr$ are given by~\cite{Aschieri:2009mc}
\begin{align}
\begin{split}
 T & \= \dd e + \tfrac12\,\big({\omega} \wedge_\star {e} + \bar{\rm R}^\alpha(\omega)\wedge_\star\bar{\rm R}_\alpha(e)\big) + \tilde\omega\wedge_\star e + e\wedge_\star\tilde\omega + \tfrac12\ast(\omega\wedge_\star\tilde e + \tilde e\wedge_\star\omega) \ , \\[4pt]
 \tilde T & \= \dd\tilde e  + \tilde\omega\wedge_\star\tilde e + \tilde e\wedge_\star\tilde\omega + \tfrac14\Tr(\omega\dwedge_\star e + e\dwedge_\star\omega) \ ,
 \end{split}
\end{align}
where $\ast$ is the Hodge duality operator on $\wedge^\bullet(\bbr^{1,2})$ induced by the Minkowski metric $\eta$. The components of $\hat R\in\Omega^2(M,\hat\frg)$ valued in $\aso(1,2)$ and $\bbr$ are
\begin{align}
\begin{split}
R & \= \dd {\omega} + \tfrac12\,[{\omega}\ds {\omega}]  + \omega\wedge_\star\tilde\omega + \tilde\omega\wedge_\star\omega \ , \\[4pt]
\tilde R & \= \dd\tilde\omega+\tilde\omega\wedge_\star\tilde\omega + \tfrac14\Tr(\omega\dwedge_\star\ast\,\omega) \ .
\end{split}
\end{align}
In the classical limit, $T$ and $R$ become the standard torsion and curvature from Section~\ref{sec:GR}, while the new fields $\tilde e$ and $\tilde\omega$ completely decouple from the classical fields with the abelian field strengths $\tilde T\=\dd\tilde e$ and $\tilde R \= \dd\tilde\omega$.

The action (without cosmological constant) which is invariant under the noncommutative local Lorentz transformations is given by~\cite{Aschieri:2009mc}
\begin{align}\label{NCAction} 
S \=  \int_M \Tr\big(\hat e\dwedge_\star \hat R\big) \= \int_M \, \varepsilon_{abc}\, e^a \wedge_\star R^{ab} + 4\,\tilde e\wedge_\star\tilde R \ .
\end{align}
The field equations resulting from the variations of the action \eqref{NCAction} are
\begin{align}\label{NCEoMII}
T\=0\=\tilde T \qquad \mbox{and} \qquad R\=0\=\tilde R \ .
\end{align}

The $L_\infty$-structure of the noncommutative local Lorentz symmetry now follows the standard rules of the classical theory from Section~\ref{sec:GR}. The underlying graded vector space 
\begin{align}
\hat V\=\hat V_0\oplus\hat V_1\oplus\hat V_2\oplus\hat V_3
\end{align}
contains the field content
\begin{align}\label{NCfieldcontent}
\begin{split}
\hat\lambda\in\hat V_0 & \=\Omega^0(M,\hat\frg) \ ,  \\[4pt]
(\hat e,\hat\omega)\in \hat V_1 & \=\Omega^1(M,\hat W)\times\Omega^1(M,\hat\frg) \ , \\[4pt]
(\hat E,\hat\varOmega)\in \hat V_2 & \= \Omega^2(M,\wedge^2(\hat W))\times\Omega^2(M,\hat W) \ , \\[4pt]
\hat\varLambda\in \hat V_3 & \= \Omega^3(M,\hat W) \ .
\end{split}
\end{align}
The differential is
\begin{align}\label{eq:ell1NCgrav}
\hat\ell_1(\hat\lambda)\=\dd\hat\lambda \ , \quad \hat\ell_1(\hat e,\hat\omega)\=(\dd\hat\omega,\dd\hat e) \qquad \mbox{and} \qquad \hat\ell_1(\hat E,\hat\varOmega)\= \dd\hat\varOmega \ ,
\end{align}
while the first few non-vanishing $2$-brackets are
\begin{align}\label{eq:ell2NCgrav}
\begin{split}
\hat\ell_2(\hat\lambda_1,\hat\lambda_2) & \= -[\hat\lambda_1 \ds \hat\lambda_2] \ , \\[4pt]
\hat\ell_2(\hat\lambda,(\hat e,\hat\omega)) & \= (-\hat\lambda\star\hat e,-[\hat\lambda \ds \hat\omega]) \ , \\[4pt]
\hat\ell_2((\hat e_1,\hat\omega_1),(\hat e_2,\hat\omega_2)) & \= - ([\hat\omega_1\ds\hat\omega_2],\hat\omega_1\wedge_\star\hat e_1 + \hat\omega_2\wedge_\star\hat e_1) \ ,
\end{split}
\end{align}
along with three others related to the field equations and Noether identities. One can check that this defines a differential graded Lie algebra, i.e. that the standard $L_\infty$-homotopy relations hold.

The noncommutative field equations \eqref{NCEoMII} can be rewritten as $\hat\calf_{(\hat e,\hat\omega)} \= 0$, where
\begin{align}
\hat\calf_{(\hat e,\hat\omega)} \= (\hat R,\hat T) \= \hat\ell_1(\hat e,\hat\omega) - \tfrac12\,\hat\ell_2((\hat e,\hat\omega),(\hat e,\hat\omega)) \ .
\end{align}
The cyclic inner product $\langle-,-\rangle:\hat V_1\otimes \hat V_2\longrightarrow\bbr$ that encodes the action \eqref{NCAction} is the generalization of \eqref{ActionII} given by
\begin{align}
\langle(\hat e,\hat \omega),(\hat E,\hat\varOmega)\rangle \= \int_M \Tr\big(\hat e\dwedge_\star\hat E + \hat\varOmega\dwedge_\star\hat\omega\big) \ .
\end{align}

Let us now study if and how the noncommutative diffeomorphism symmetry can be incorporated into this classical $L_\infty$-structure. For a vector field $\xi\in\Gamma(TM)$, the action of noncommutative diffeomorphisms on the various fields \eqref{NCfieldcontent} of three-dimensional noncommutative gravity is via the braided Lie derivative \eqref{StarLieDer}:
\begin{align}
\hat\delta_\xi \hat{\mbf T} \= \CL_\xi^\star\hat{\mbf T} \ .
\end{align}
Indeed, the action \eqref{NCAction} is invariant under braided diffeomorphisms~\cite{Aschieri:2009mc}.

The dynamical part of the $L_\infty$-algebra is not altered by the inclusion of diffeomorphisms, only its symmetry part, which we modify by extending the vector spaces $\hat V_0$ and $\hat V_3$ respectively by
\begin{align}
\xi\in\Gamma(TM)\subset \hat V_0 \qquad \mbox{and} \qquad \varXi\in\Omega^1(M,\Omega^3(M))\subset\hat V_3 \ ,
\end{align}
as in the classical case. The brackets \eqref{eq:ell1NCgrav} and  \eqref{eq:ell2NCgrav} are correspondingly supplemented with
\begin{align}
\hat \ell_1(\xi)\=(0,0) \quad , \quad \hat\ell_2(\xi_1,\xi_2)\=[\xi_1,\xi_2]^\star \quad \mbox{,} \quad \hat\ell_2(\xi,(\hat e,\hat \omega))\=(\CL_\xi^\star\hat e,\CL_\xi^\star\hat\omega) \ .
\end{align}
Of course, $\hat\ell_2(\xi_1,\xi_2)$ is the bracket of the braided Lie algebra $\Gamma_\star(TM)$ of noncommutative vector fields on $M$, and so is not antisymmetric, but rather braided antisymmetric:
\begin{align}
\hat\ell_2(\xi_1,\xi_2)\=-\hat \ell_2(\bar{\rm R}^\alpha(\xi_2),\bar{\rm R}_\alpha(\xi_1)) \ .
\end{align}
As in the case of the noncommutative scalar field theory from Section~\ref{sec:braided}, we can antisymmetrize the bracket by hand and define
\begin{align}\label{AntisymmProposition}
\hat\ell_2^{\,{\rm sym}}(\xi_1,\xi_2)\=\tfrac12\,\big([\xi_1,\xi_2]^\star-[\xi_2,\xi_1]^\star\big) \ .
\end{align}

However, while most of the homotopy relations hold, there is one  relation which is violated:
\begin{align}\label{JacobiProblem1}
\hat\ell_2^{\,{\rm sym}}(\hat\ell_2^{\,{\rm sym}}(\xi_1,\xi_2),\xi_3) + \hat\ell_2^{\,{\rm sym}}(\hat\ell_2^{\,{\rm sym}}(\xi_3,\xi_1),\xi_2) + \hat\ell_2^{\,{\rm sym}}(\hat\ell_2^{\,{\rm sym}}(\xi_2,\xi_3),\xi_1) \ \neq \ 0 \ .
\end{align}
The vanishing of this expression would be the Jacobi identity for the antisymmetrized bracket \eqref{AntisymmProposition}, which is violated because of the \emph{braided} Jacobi identity \eqref{StarBracket2} for the braided Lie algebra $\Gamma_\star(TM)$. Generally, the right-hand side of the homotopy Jacobi identity \eqref{JacobiProblem1} can be equal to $-\hat\ell_1(\hat\ell_3(\xi_1,\xi_2,\xi_3))$, which requires extending the cochain complex $(\hat V,\hat\ell_1)$ by a homogeneous subspace $\hat V_{-1}$ in degree~$-1$ (and dually by $\hat V_4$ in degree~$4$) in order to accomodate a non-zero $3$-bracket $\hat\ell_3(\xi_1,\xi_2,\xi_3)$. 

This suggests that, from the perspective of standard symmetries and their $L_\infty$-structure, the twisted diffeomorphism symmetry of noncommutative gravity may provide a reducible or \emph{higher} gauge symmetry. While this is a tantalizing feature, it is not clear to what extent this approach is technically feasible, as it may require the introduction of infinitely many higher brackets, as in the $L_\infty$-algebra approaches to generic noncommutative gauge theories~\cite{Blumenhagen:2018kwq,Kupriyanov:2019ezf,Kupriyanov:2021cws,Abla:2022wfz}, however here with additional corresponding homogeneous subspaces in negative degrees. Moreover, the physical meaning of such an extension is not immediately apparent, as it vanishes in the classical limit. We conclude that the twisted diffeomorphism symmetry does not fit naturally (if at all) into the standard $L_\infty$-algebra picture.

\paragraph{Braided $\mbf{L_\infty}$-structure.}
Motivated by the issues encountered above, we deform the $L_\infty$-structure itself to make it compatible with the twisted diffeomorphisms, leading to our novel mathematical notion of a braided $L_\infty$-algebra and the corresponding noncommutative field theories with braided gauge symmetries. In contrast to the approach reviewed above, the transformations \eqref{NCLorTransf} are now replaced with the braided Lorentz transformations
\begin{align}
\delta_\lambda^\star e \= -\lambda\star e \qquad \mbox{and} \qquad \delta_\lambda^\star\omega \= \dd\omega-[\lambda,\omega]^\star \= \dd\lambda-\big(\lambda\star \omega - \bar{\rm R}^\alpha(\omega)\star\bar{\rm R}_\alpha(\lambda)\big) \ .
\end{align}
These transformations close in the Lie algebra, and there is no need to centrally extend the Lorentz algebra (or more generally to extend to the universal enveloping algebra). No new degrees of freedom are introduced, and the noncommutative theory reduces exactly to the standard commutative theory in the classical limit.

The braided noncommutative deformation of three-dimensional ECP gravity follows much the same route as the braided Chern-Simons theory from Section~\ref{sec:braided}, resulting in the `naive' deformed gravity theory obtained by replacing products of fields with star-products in the classical theory; this appears to be a general feature of the braided deformation of classical field theories whose $L_\infty$-algebra is a differential graded Lie algebra. See~\cite{Ciric:2020eab} for a brief review of the $d=3$ theory and~\cite{Ciric:2021rhi} for a more detailed account. In $d\=4$ dimensions, a genuinely new noncommutative theory of gravity arises, similarly to the braided noncommutative Yang-Mills theory from Section~\ref{sec:braided}. Let us summarise the main features; further details can be found in~\cite{Ciric:2021rhi}.

We add a cosmological constant $\Lambda$. As usual the differential $\ell_1^\star\=\ell_1$ from Section~\ref{sec:GR} is unchanged and thus given by \eqref{eq:ell14d}. The $2$-brackets are twisted to
\begin{align}\label{eq:twisted2brackets}
\begin{split}
\ell_2^\star((\xi_1,\lambda_1),(\xi_2,\lambda_2))&\=([\xi_1,\xi_2]^\star,-[\lambda_1,\lambda_2]^\star+\CL^\star_{\xi_1}\lambda_2-\CL^\star_{\bar{\rm R}^\alpha(\xi_2)}\bar{\rm R}_\alpha(\lambda_1)) \ , \\[4pt]
\ell^\star_2((\xi,\lambda),(e,\omega))&\=(-\lambda\star e+\CL^\star_\xi e,-[\lambda,\omega]^\star+\CL^\star_\xi\omega) \ , \\[4pt]
\ell^\star_{2}((e_{1},\omega_{1}),(e_{2},\omega_{2}))&\=-(e_{1} \dwedge_\star \dd \omega_{2}
+ e_{2} \dwedge_\star \dd \omega_{1} , e_{1}
\dwedge_\star \dd e_{2} + e_{2} \dwedge_\star \dd e_{1}) \ ,
\end{split}
\end{align}
and again with three more $2$-brackets related to the field equations and Noether identities. Finally, the remaining non-zero braided $3$-bracket is given by 
\begin{align}
\small
\begin{split}
& \ell^\star_{3}((e_{1},\omega_{1}),(e_{2},\omega_{2}),(e_{3},\omega_{3})) \\[4pt]
& \hspace{0.5cm} \=-( e_{1}\dwedge_\star[\omega_{2} , \omega_{3}]^\star + \bar{\rm R}^\alpha(e_{2}) \dwedge_\star [\bar{\rm R}_\alpha(\omega_{1}) , \omega_{3}]^\star  + [\omega_{1} , \omega_{2}]^\star\dwedge_\star e_3+6\,\Lambda\,e_1\dwedge_\star e_2\dwedge_\star e_3, \\[2pt]
& \hspace{1.5cm} \ e_{1} \dwedge_\star (\omega_{2} \wedge_\star e_{3}) 
+e_1\dwedge_\star(\bar{\rm R}^\alpha(\omega_3)\wedge_\star\bar{\rm R}_\alpha(e_2)) \\ & \hspace{2cm} \ + \bar{\rm R}^\alpha(e_{2})\dwedge_\star (\bar{\rm R}_\alpha(\omega_{1}) \wedge_\star e_{3}) +\bar{\rm R}^\alpha(e_{2})\dwedge_\star (\bar{\rm R}^\beta(\omega_{3}) \wedge_\star \bar{\rm R}_\beta\,\bar{\rm R}_\alpha(e_{1})) \\ & \hspace{2.5cm} \ -  (\omega_{1} \wedge_\star e_{2})\dwedge_\star e_3 -  (\bar{\rm R}^\alpha(\omega_{2}) \wedge_\star \bar{\rm R}_\alpha(e_{1}))\dwedge_\star e_3 ) \ .
\end{split}
\normalsize
\end{align}

The braided gauge transformations
\begin{align}
\delta_{(\xi,\lambda)}^\star(e,\omega)\=\ell_1^\star(\xi,\lambda) + \ell_2^\star((\xi,\lambda),(e,\omega)) \= (-\lambda\star e+\CL^\star_\xi e,\dd\lambda-[\lambda,\omega]^\star+\CL_\xi^\star\omega)
\end{align}
close a braided Lie algebra with braided Lie bracket $-\ell_2^\star$ of gauge parameters given in \eqref{eq:twisted2brackets}. It follows that braided gravity is 
{invariant} under the braided semi-direct product
\begin{align}
\Gamma_\star(TM)\ltimes_\star\Omega^0_\star(M,\aso(1,3))
\end{align}
of the braided Lie algebra $\Gamma_\star(TM)$ of vector fields on $M$ and the braided Lie algebra $\Omega^0_\star(M,\frg)$ from Section~\ref{sec:braided} with $\frg\=\aso(1,3)$, $W\=\bbr^{1,3}$ and $p\=1$. Thus our braided theory of noncommutative gravity preserves the semi-direct product structure of the classical symmetry algebra \eqref{eq:classsemidirect}.

The field equations $\calf_{(e,\omega)}^\star\=(0,0)$ of braided ECP gravity in four dimensions are given by
\begin{align}
\calf_{(e,\omega)}^\star\=-\tfrac12\,\ell_2^\star((e,\omega),(e,\omega)) - \tfrac16\,\ell_3^\star((e,\omega),(e,\omega),(e,\omega))\=(\calf_e^\star,\calf_\omega^\star) \ .
\end{align}
Defining the braided left/right torsion and curvature of the spin connection $\omega$ as
\begin{align}
T_{\lact,\ract}^\star \ := \ \dd^\omega_{\star\lact,\ract}e \qquad \mbox{and} \qquad R^\star \ := \ \dd\omega+\tfrac12\,[\omega,\omega]^\star \ ,
\end{align}
they can be written as
\begin{align}
\begin{split}
& e\dwedge_\star T_\lact^\star - T^\star_\ract\dwedge_\star e -\,\dd^\omega_{\star\lact}(e\dwedge_\star e) - \dd^\omega_{\star\ract}(e\dwedge_\star e) \=0 \ , \\[4pt]
& 2\,e\dwedge_\star R^\star + 2\,R^\star\dwedge_\star e + 6\,\Lambda\,e\dwedge_\star e\dwedge_\star e \\ & \qquad +\, e\dwedge_\star\dd\omega + \dd\omega\dwedge_\star e + \bar{\rm R}^\alpha(e)\dwedge_\star[\bar{\rm R}_\alpha(\omega),\omega]^\star \=0 \ .
\end{split}
\end{align}
Despite the appearance of individually non-covariant terms, taken together these equations are covariant, as guaranteed on general grounds by the braided $L_\infty$-structure:
\begin{align}
\delta^\star_{(\xi,\lambda)}\calf^\star_{(e,\omega)} \= \ell_2^\star((\xi,\lambda),(\calf_e^\star,\calf_\omega^\star)) \ ,
\end{align}
which leads to
\begin{align}
\delta_{(\xi,\lambda)}^\star\calf_e^\star\=-\lambda\star\calf_e^\star + \CL_\xi^\star\calf_e^\star \qquad \mbox{and} \qquad \delta_{(\xi,\lambda)}^\star\calf_\omega^\star\=-\lambda\star\calf_\omega^\star + \CL_\xi^\star\calf_\omega^\star \ .
\end{align}
In the classical limit they reduce respectively to the torsion-free condition for the spin connection and the four-dimensional vacuum Einstein equations with cosmological constant.

The action follows from twisting the classical ECP inner product \eqref{ActionII} to
\begin{align}
\langle(e,\omega),(E,\varOmega)\rangle_\star\=\int_M \Tr(e\dwedge_\star E - \omega\dwedge_\star\varOmega)
\end{align}
on $V_1\otimes V_2$. 
Then the action
\begin{align}
\begin{split}
S^\star&\=\tfrac12\,\langle(e,\omega),\ell_1^\star(e,\omega)\rangle_\star - \tfrac16\,\langle(e,\omega),\ell_2^\star((e,\omega),(e,\omega))\rangle_\star \\ & \qquad - \tfrac1{24} \, \langle(e,\omega),\ell_3^\star((e,\omega),(e,\omega),(e,\omega))\rangle_\star
\end{split}
\end{align}
leads to
\begin{align}
\begin{split}
S^\star & \= \int_M \Tr\Big(\frac12\,e\dwedge_\star e\dwedge_\star R^\star+\frac\Lambda4\, e\dwedge_\star e\dwedge_\star e\dwedge_\star e \Big) \\ & \qquad -\,\frac1{24}\,\int_M \Tr\Big(\omega\dwedge_\star\big(2\,e\dwedge_\star T_\lact^\star - 2\,T^\star_\ract\dwedge_\star e + \dd_{\star\lact}^\omega(e\dwedge_\star e) + \dd_{\star\ract}^\omega(e\dwedge_\star e)\big)\Big) \ .
\end{split}
\end{align}
Again, despite the appearance of the second line, this is gauge invariant with a good classical limit that does not involve spurious degrees of freedom. The associated Noether identities in this case are quite complicated, and are discussed in detail in~\cite{Ciric:2021rhi}. This theory defines a new deformation of general relativity.

\acknowledgments

The author warmly thanks Marija Dimitrijevi\'c \'Ciri\'c, Grigorios Giotopoulos, Hans Nguyen, Voja Radovanovi\'c and Alexander Schenkel for collaborations and discussions over the past few years, upon which this contribution is based. He would also like to thank the organisors of the Corfu Summer Institute 2021 and the Humboldt Kolleg on ``Quantum Gravity and Fundamental Interactions'' for the stimulating meeting and the invitation to deliver a talk.
This work was supported by
the Consolidated Grant ST/P000363/1 
from the UK Science and Technology Facilities Council.

\end{document}